\documentclass[apj]{emulateapj}

\usepackage{apjfonts}
\shorttitle{SSDI with a focal plane holographic diffuser}
\shortauthors{Lafreni\`{e}re et al.}

\newcommand{\drm}{\mathrm{d}}

\begin{document}

\title{Improving the speckle noise attenuation of simultaneous spectral differential imaging with a focal plane holographic diffuser}

\author{David Lafreni\`ere\altaffilmark{A}, Ren\'e Doyon\altaffilmark{A}, Daniel Nadeau\altaffilmark{A}, \'Etienne Artigau\altaffilmark{A,B}, Christian Marois\altaffilmark{C}, and Mathilde Beaulieu\altaffilmark{a}}
\altaffiltext{A}{D\'epartement de physique and Observatoire du Mont M\'egantic, Universit\'e de Montr\'eal, C.P. 6128, Succ. Centre-Ville, Montr\'eal, QC, Canada H3C 3J7}
\altaffiltext{B}{Gemini Observatory, Southern Operations Center, Association of Universities for Research in Astronomy, Inc., Casilla 603, La Serena, Chile}
\altaffiltext{C}{Institute of Geophysics and Planetary Physics L-413, Lawrence Livermore National Laboratory, 7000 East Ave, Livermore, CA 94550}
\email{david@astro.umontreal.ca doyon@astro.umontreal.ca \newline nadeau@astro.umontreal.ca eartigau@gemini.edu \newline cmarois@igpp.ucllnl.org mathilde@astro.umontreal.ca} 

\begin{abstract}

Direct exoplanet detection is limited by speckle noise in the point spread function (PSF) of the central star. This noise can be reduced by subtracting PSF images obtained simultaneously in adjacent narrow spectral bands using a multi-channel camera (MCC), but only to a limit imposed by differential optical aberrations in the MCC. To alleviate this problem, we suggest the introduction of a holographic diffuser at the focal plane of the MCC to convert the PSF image into an incoherent illumination scene that is then re-imaged with the MCC. The re-imaging is equivalent to a convolution of the scene with the PSF of each spectral channel of the camera. Optical aberrations in the MCC affect only the convolution kernel of each channel and not the PSF globally, resulting in better correlated images. We report laboratory measurements with a dual channel prototype ($1.575\ \mu$m and $1.625\ \mu$m) to validate this approach. A speckle noise suppression factor of 12-14 was achieved, an improvement by a factor $\sim$5 over that obtained without the holographic diffuser. Simulations of realistic exoplanet populations for three representative target samples show that the increase in speckle noise attenuation achieved in the laboratory would roughly double the number of planets that could be detected with current adaptive optics systems on 8-m telescopes.

\end{abstract} 
\keywords{Instrumentation: adaptive optics --- planetary systems --- stars: imaging --- techniques: image processing ---  techniques: high angular resolution}

\section{Introduction}

Direct detections of faint exoplanets near bright stars are essential to extend the census of planetary companions to separations that are beyond the reach of radial velocity surveys and to enable measurements of exoplanets' physical properties through follow-up multi-color photometry and spectroscopy. This difficult endeavour, already being tackled by many, is one of the major goals for next generation instruments on 8-m class telescopes and future 30- to 100-m telescopes. The main difficulty stems from imperfections in the optics that produce bright quasi-static speckles in the point spread function (PSF) of the central star \citep{trident, biller04, masciadri05}. The subtraction of reference PSF images is a very efficient way of reducing this speckle noise. A good example is angular differential imaging (ADI, \citealp{maroisADI, lafreniereADI}), in which images of the target, obtained while the field of view rotates, are used for subtracting the stellar PSF. This technique is among the most successful at suppressing speckle noise for ground-based imaging; however, it is inefficient at small angular separation ($\lesssim1^{\prime\prime}$) because the time required for sufficient natural rotation of the field of view is too long.

Simultaneous spectral differential imaging (SSDI) is a PSF subtraction technique that is efficient at all angular separations. It consists in the simultaneous acquisition of images in adjacent narrow spectral bands within a spectral range where the stellar and planetary spectra differ appreciably \citep{racine99, marois2000, smith87, rosenthal96}. Judicious image combination and subtraction removes the stellar PSF and leaves that of any companion. Note that because the PSF images are acquired in different bandpasses, they must be rescaled prior to subtraction (diffraction scales as $\lambda$). At larger separations, this scaling allows the technique to work even in the absence of differential spectral features since a companion would not overlie itself in the rescaled images.

The standard implementation of SSDI is done with a multi-channel camera (MCC). The speckle noise suppression achievable with such instruments is hampered by differential aberrations between the spectral channels \citep{trident}. For the triple imager TRIDENT (\citealp{trident}, Canada-France-Hawaii Telescope) and the quadruple imager SDI (\citealp{sdi}, Very Large Telescope), the subtraction of two images obtained simultaneously through different channels yields a speckle noise attenuation of only $\sim$2-2.5 (\citealp{trident} and Appendix~\ref{sect:appendix}). Strategies such as multiple rotations of the instrument or observation of reference stars \citep{trident, biller04} have been used to attenuate further the residuals left by the incomplete simultaneous multi-channel subtraction, and 5$\sigma$ contrast limits for magnitude differences of $\sim$9 and $\gtrsim$10 have been achieved at an angular separation of 0\farcs5 with TRIDENT and SDI respectively \citep{trident, biller06}. However, the approaches used to circumvent the problem of differential aberrations often complicate data acquisition and decrease the observing efficiency at the telescope. As several MCCs are currently being developed by major observatories for future exoplanet searches, e.g. NICI \citep{nici} for Gemini South, SPHERE/IRDIS \citep{sphere} for the VLT, or HiCIAO \citep{hiciao1,hiciao2} for Subaru, the development of techniques to improve the PSF subtraction performance of MCCs is of great interest.

In this paper, we propose to minimize the effect of differential aberrations by introducing a holographic diffuser (HD) at the focal plane of an MCC (see also \citealp{diff_spie}). This simple approach provides a direct gain in speckle attenuation without the need to modify the observing strategy. The concept and performance estimates are presented in \S\ref{sect:concept}. Measurements and results using a laboratory prototype are reported in \S\ref{sect:results}, and \S\ref{sect:perf} presents estimates of the detection limits that such a device should reach on 8-m telescopes using current adaptive optics (AO) systems, along with an assessment of the corresponding exoplanet detection efficiency. Improvements in performance for a dedicated camera, implications of the laboratory results for other types of SSDI instruments, and considerations for implementation of this concept in an existing camera are discussed in \S\ref{sect:discussion}.

\section{Multi-channel imager with holographic diffuser}\label{sect:concept}

\subsection{Concept}\label{sect:principle}
A light shaping HD is a random surface relief hologram that diffuses the light incident upon it into a controlled angular distribution. When a coherent wavefront goes through an ideal HD, large phase errors are introduced over small scales and spatial coherence is lost; thus, every point of the wavefront effectively becomes an independent source emitting with an angular distribution controlled by the HD. Hence, an HD located at a focal plane converts the PSF image into an incoherent illumination scene, which can then be re-imaged by an MCC. This is equivalent to a convolution of the scene with the PSF of each channel of the MCC. Optical aberrations in the MCC affect only the convolution kernel of each channel rather than the PSF globally; the effect of differential aberrations is merely a convolution of the same PSF image with different kernels. Hence, contrary to standard MCC imaging, the speckle pattern remains the same in all channels and it is only the light distribution within each speckle that may vary slightly depending on the size of the convolution kernel and on the amount of aberrations present in the MCC. If the size of the convolution kernel is smaller than the size of a speckle in the incident PSF image, then most of the light will remain inside the same speckle after convolution and the correlation between the different channels will be high.

An HD is a better choice than other diffuser types for use with an MCC because its diffusing cone angle can be made small, allowing efficient coupling to relatively large $f/\#$ ($\sim$8-32), typical of existing cameras. Moreover, the diffusing properties of an HD are wavelength independent. It should be noted that since the diffuser breaks the coherence of the wavefront, it precludes the use of a coronagraph downstream from it. However, the diffuser poses no problem for a coronagraph placed in front of it.

\begin{figure}
\epsscale{1.}
\plotone{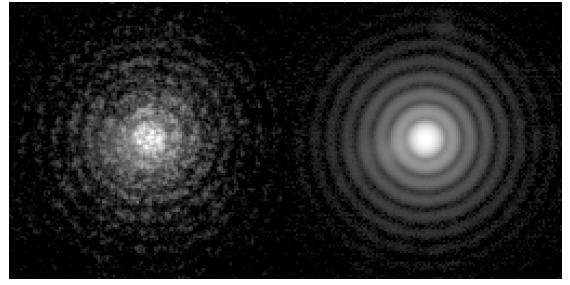}
\caption{\label{fig:diffrot} Images of a PSF obtained with a camera equipped with an HD at its entrance focal plane. The HD was fixed ({\it left}) or moving continuously ({\it right}) during the exposure; see \S\ref{sect:principle} for more detail. The images are shown with a logarithmic intensity scale. They were acquired using the experimental setup described in \S\ref{sect:setup}. }
\end{figure}

A real HD does not behave exactly as described above because the size of the structures on its surface is not nnegligible compared to the full-width-at-half-maximum (FWHM) of the incident PSF. The surface of an HD can be thought of as a random collection of microlenses, each refracting light slightly differently such that the overall diffusion of light results in the desired angular distribution. A real HD thus transforms the wavefront incident upon it into a random collection of sources, or micro-pupils, of finite size rather than ideal, infinitesimally spaced, point sources. Accordingly, the image recorded by an MCC equipped with a focal plane HD would correspond to the smooth illumination scene incident upon the HD broken up into many tiny spots, whose position and intensity are determined by the position and amount of light intercepted by the corresponding microlenses (see Figure~\ref{fig:diffrot}). As the micro-pupils produced by the microlenses are usually not resolved by the MCC, the shape, or FWHM, of each tiny spot is determined by the focal ratio of the MCC. This breakup of the smooth illumination scene into a collection of tiny spots is problematic, but a simple solution exists. If the HD is moved slightly in a plane tangent to its surface, then each tiny spot will move on the detector by a corresponding amount, and the intensity of each spot will change according to the new amount of light intercepted by the corresponding microlens. In the limit where the HD moves continuously during an exposure, the detector is smoothly illuminated and, effectively, the image recorded by the MCC corresponds to a convolution of the illumination scene incident upon the HD with the PSF of the MCC (see Figure~\ref{fig:diffrot}).

The imaging process described above can be understood mathematically as follows. If $I_1$ and $I_2$ represent the PSFs of the two channels of a dual imager for a point source located at its entrance focal plane, and $I_{\rm inc}$ represents the PSF incident upon the HD, then the resulting image, in each channel, is given by the convolution $I_{\rm inc}\star I_1$ or $I_{\rm inc}\star I_2$. The residual image $R$ obtained by subtracting the images of both channels is then given by

\begin{equation}\label{eq:res}
R=(I_{\rm inc}\star I_1)-(I_{\rm inc}\star I_2)=I_{\rm inc}\star (I_1-I_2).
\end{equation}

\noindent The residual image can thus be viewed as being equal to the difference of the two MCC PSFs, convolved with the PSF incident upon the HD. As the spatial scale of the residual noise in $(I_1-I_2)$ is the FWHM of $I_1$ and $I_2$, denoted $\theta_{\rm MCC}$, and the width of the convolution kernel is the FWHM of $I_{\rm inc}$, denoted $\theta_{\rm inc}$, the above convolution should reduce the amplitude of the residual noise in $(I_1-I_2)$ by a factor dependent upon the ratio $\theta_{\rm inc}/\theta_{\rm MCC}$. This ratio represents the number of characteristic lengths over which the residuals are averaged by the convolution. A larger ratio $\theta_{\rm inc}/\theta_{\rm MCC}$ should lead to a larger attenuation of the noise.

It was demonstrated in \citet{trident} that the noise attenuation achieved by a dual imager is given by $\sigma_{\rm MCC}/\Delta\sigma_{\rm MCC}$, where $\sigma_{\rm MCC}$ is the amount of aberration in each channel and $\Delta\sigma_{\rm MCC}$ is the amount of differential aberration. Thus when using an HD one can expect, based on Eq.~(\ref{eq:res}), that the residual noise also scales as $\sigma_{\rm MCC}/\Delta\sigma_{\rm MCC}$, but with an additional factor due to the convolution. While one should still minimize differential aberrations to maximize the attenuation, important leverage is introduced through the ratio $\theta_{\rm inc}/\theta_{\rm MCC}$, or equivalently through $(f/\#)_{\rm inc}/(f/\#)_{\rm MCC}$, where $(f/\#)_{\rm inc}$ and $(f/\#)_{\rm MCC}$ are the focal ratios of the beam incident upon the HD and of the MCC. For better performance, these ratios should be made as large as possible. This would also minimize the loss in spatial resolution resulting from the convolution of $I_{\rm inc}$ with $I_1$ or $I_2$. From convolutions of Airy functions, it can be shown that the loss in resolution is $\sim$12\% for $(f/\#)_{\rm inc}/(f/\#)_{\rm MCC}=2$ and less than 5\% for $(f/\#)_{\rm inc}/(f/\#)_{\rm MCC}\ge$3.

\subsection{Speckle noise attenuation estimation}

Numerical simulations were done to estimate the speckle noise attenuation that may be achieved with a dual channel camera equipped with an HD. 
For each simulation, three PSFs were created: the PSF incident upon the HD ($I_{\rm inc}$) and the PSFs of both channels of the MCC ($I_1$ and $I_2$). Arrays of 2048$\times$2048 pixels were used for all calculations. The PSFs were obtained as the square modulus of the Fourier Transform of the complex pupil function. The MCC PSFs FWHM, $\theta_{\rm MCC}$, was fixed to 5 pixels, and the incident PSF FWHM, $\theta_{\rm inc}$, was varied to test the effect of the ratio $\theta_{\rm inc}/\theta_{\rm MCC}$. Wavefront aberrations having a power-law power spectrum of index -2.7, a value appropriate for AO systems and astronomical cameras \citep{trident},  were included in the simulations. Different values of the wavefront error of the incident PSF $\sigma_{\rm inc}$, of $\sigma_{\rm MCC}$, and of $\Delta\sigma_{\rm MCC}$ were tried to see their effect.

\begin{figure}
\epsscale{1.}
\plotone{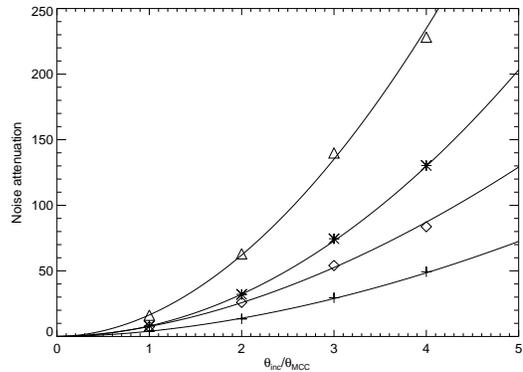}
\caption{\label{fig:attsim} Speckle noise attenuation from numerical simulations as a function of $\theta_{\rm inc}/\theta_{\rm MCC}$, the ratio of the FWHM of the PSF incident on the HD over that of the intrinsic MCC PSF. The values of $\sigma_{\rm inc}$ and $\sigma_{\rm MCC}$, in nm, are respectively 80 and 50 ({\it crosses}), 80 and 25 ({\it stars}), 120 and 50 ({\it diamonds}), and 120 and 25 ({\it triangles}). All results are for $\sigma_{\rm MCC}/\Delta\sigma_{\rm MCC}=2$. The curves are the fit of Eq.~(\ref{eq:attsim}) to each set of points.}
\end{figure}

Once the three PSFs have been obtained, $I_1$ and $I_2$ are normalized to a sum of unity and are subtracted from each other. This difference is then convolved with $I_{\rm inc}$ to obtain the residual image. The noise attenuation is finally determined. The PSF noise is defined as the standard deviation over an annulus of the PSF image after subtraction of an azimuthally averaged profile. The noise attenuation $\eta$ is defined as the ratio of the PSF noise in one channel over the noise in the residual image. Many values of $\theta_{\rm inc}/\theta_{\rm MCC}$, $\sigma_{\rm inc}$, $\sigma_{\rm MCC}$, and $\Delta\sigma_{\rm MCC}$ were tried; the results are presented in Table~\ref{tbl:attsim} and Figure~\ref{fig:attsim}. We note that the total amount of static aberration for TRIDENT at the CFHT was estimated at 130~nm, of which 50~nm originated from the camera, and 120~nm originated from the AO system and telescope \citep{trident}; the values of $\sigma_{\rm inc}$ and $\sigma_{\rm MCC}$ tested here are therefore representative of real systems. The results indicate that the noise attenuation increases with $\sigma_{\rm MCC}/\Delta\sigma_{\rm MCC}$ and with $\theta_{\rm inc}/\theta_{\rm MCC}$, in agreement with the previous discussion. 

\begin{deluxetable}{ccccccccc}
\tablecolumns{9}
\tablewidth{240pt}
\tablecaption{Speckle noise attenuation from numerical simulations. \label{tbl:attsim}}
\tablehead{
\multicolumn{3}{c}{Optical aberrations} & \multicolumn{4}{c}{Att. for $\theta_{\rm inc}/\theta_{\rm MCC}$ equal to} & \multicolumn{2}{c}{Fit of Eq.~(\ref{eq:attsim})} \\
\colhead{$\sigma_{\rm inc}$} & \colhead{$\sigma_{\rm MCC}$} & \colhead{$\frac{\sigma_{\rm MCC}}{\Delta\sigma_{\rm MCC}}$} & 
\colhead{1} & \colhead{2} & \colhead{3} & \colhead{4} & \colhead{$\alpha$} & \colhead{$\beta$}
}
\startdata
 80 & 50 & 1 &   2.1 &   6.9 &  15.2 &  25.3 &  2.1 & 1.80 \\
 80 & 50 & 2 &   4.1 &  13.4 &  29.5 &  49.3 &  2.0 & 1.80 \\
 80 & 50 & 3 &   6.1 &  20.0 &  44.2 &  73.8 &  2.0 & 1.81 \\
\tableline
 80 & 25 & 1 &   4.1 &  16.2 &  37.6 &  65.7 &  4.1 & 2.01 \\
 80 & 25 & 2 &   8.1 &  32.1 &  74.5 & 130.3 &  4.0 & 2.01 \\
 80 & 25 & 3 &  12.2 &  48.0 & 111.6 & 195.2 &  4.0 & 2.01 \\
\tableline
120 & 50 & 1 &   3.8 &  13.4 &  27.7 &  42.9 &  3.9 & 1.76 \\
120 & 50 & 2 &   7.4 &  26.0 &  54.1 &  83.7 &  3.8 & 1.76 \\
120 & 50 & 3 &  11.0 &  38.8 &  81.0 & 125.8 &  3.7 & 1.77 \\
\tableline
120 & 25 & 1 &   8.2 &  31.9 &  70.7 & 115.5 &  8.3 & 1.92 \\
120 & 25 & 2 &  16.2 &  63.0 & 139.9 & 228.2 &  8.2 & 1.92 \\
120 & 25 & 3 &  24.3 &  94.3 & 209.4 & 342.3 &  8.2 & 1.92
\enddata
\end{deluxetable}

Using a dual channel camera, the maximum value of the attenuation, limited by chromaticity of the PSF, is equal to $\lambda/\Delta\lambda$ \citep{marois2000, trident}, where $\lambda$ is the wavelength of the first channel and $\Delta\lambda$ is the difference in wavelength of the two channels. The maximum attenuation is thus $\sim$30 for bandpasses located just outside and just inside the methane absorption feature in the spectrum of cold dwarfs at $\sim$1.6~$\mu$m (e.g. $\lambda_1\sim1.575\mu$m and $\lambda_2\sim1.625\mu$m). From Table~\ref{tbl:attsim}, the attenuation should be limited by chromaticity of the PSFs rather than by differential aberrations for $(f/\#)_{\rm inc}/(f/\#)_{\rm MCC}\gtrsim$3-4.

It is found that the attenuation is very well approximated by the function

\begin{equation}\label{eq:attsim}
\eta=\alpha \left(\frac{\sigma_{\rm MCC}}{\Delta\sigma_{\rm MCC}}\right) \left( \frac{\theta_{\rm inc}}{\theta_{\rm MCC}} \right)^\beta,
\end{equation}

\noindent where $\alpha$ and $\beta$ are parameters that depend upon $\sigma_{\rm inc}$ and $\sigma_{\rm MCC}$ (see Figure~\ref{fig:attsim}); their values are indicated in Table~\ref{tbl:attsim}. Thus in a regime representative of real AO systems and MCC cameras, the attenuation that may be achieved with a system equipped with an HD roughly scales as the square of $\theta_{\rm inc}/\theta_{\rm MCC}$.

\section{Experimental results}\label{sect:results}

\subsection{Dual imager testbed}\label{sect:setup}

A simple testbed operating in the 1.55-1.65~$\mu$m wavelength range was assembled on an optical bench to validate the concept presented in the last section. The experimental setup consists of a PSF generating system followed by a two-channel camera.

The PSF generating imaging system is composed of a bi-convex and a plano-convex lens (Thorlabs, LB1056 and LA1172) and a 6.3~mm diameter pupil stop, which is located immediately before the first lens. The system is telecentric and has an exit focal ratio of $(f/\#)_{\rm inc}=64$. This focal ratio was chosen based on the results of the previous section and the $f/16$ focal ratio of the dual imager described below. The PSF is generated by imaging a 25~$\mu$m pinhole illuminated from the back by a white light source. Due to the small number of optical components in this system and to the small dimension of the beam, the Strehl ratio achieved is very high and no speckles are visible in the resulting PSF image. This is problematic for testing a technique aimed specifically at reducing speckle noise; it was thus necessary to introduce more aberrations in the system. This was achieved by placing an $H$-band filter just before the pupil of the system, resulting in a Strehl ratio of $\sim$92\%, or $\sim$75 nm rms of aberrations. 

\begin{figure*}
\epsscale{1.}
\plotone{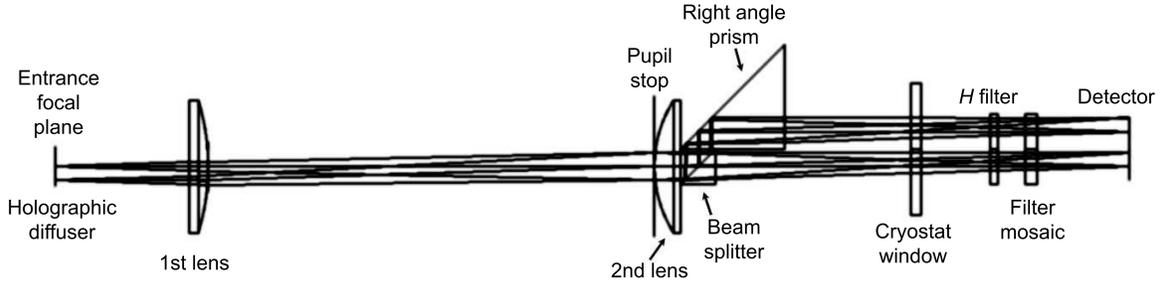}
\caption{\label{fig:layout} Optical layout of the dual imager testbed. Rays are drawn from two field positions at the entrance focal plane. A separate optical system (not shown) is placed in front of the dual imager to produce a PSF on the surface of the HD. See text for more detail.}
\end{figure*}

The exit focal plane of the PSF generating system corresponds to the entrance focal plane of the two-channel camera. An HD can be put in/taken out of the light path at this precise location without altering any other part of the setup. For reasons explained in \S\ref{sect:concept}, the HD is mounted inside a ball bearing coupled to a small electric motor by a gear belt and spins at a rate of $\sim$10~Hz. The rotation axis (center) of the HD is offset by $\sim$6~mm from the optical axis to maximize the effective area of the HD seen by the dual imager. The rotation rate was chosen to ensure that at least a few complete rotations would occur during a single exposure.

A diagram of the two-channel camera is shown in Figure~\ref{fig:layout}. The camera consists primarily of two plano-convex lenses (CVI Laser, PLCX-38.1-64.4-C and JML Optical Industries, PPX12925/000); an 8~mm pupil stop is located immediately before the second lens. This camera has an entrance focal ratio of $(f/\#)_{\rm MCC}=16$; this is roughly the smallest focal ratio that could be realized easily using simple commercially available lenses. A magnification of unity was chosen for simplicity. Beam splitting is made directly after the second lens with a 50/50 beamsplitter cube (Thorlabs, BS012). The beam reflected at 90 degrees is folded back toward the detector by mean of a right angle prism (Bernhard Halle, UPB0.30); the side dimension of this prism was selected such that the extra path length in glass compensates precisely the increased physical path length of this channel, allowing both channels to reach focus in the same plane. The detector used is a Hawaii-1 (Rockwell Scientific Company), it is housed inside an IRLab cryostat. A square filter mosaic consisting of two abutted rectangular filters of bandwidth 3\%, centered on 1.575~$\mu$m and 1.625~$\mu$m respectively, is located in front of the detector and allows imaging through different bandpasses in the two channels. An additional $H$-band filter is placed inside the cryostat to block radiation longward of 1.8 $\mu$m as the narrow band filters are not blocked outside of the $H$ band. This dual imager has a square field of view 4~mm on a side and a Strehl ratio $>95$\% over the entire field of view in each channel, as estimated from Zemax. The FWHM of the PSF at the detector is 5.5 pixels. There is no anti-reflection coating on any of the optical components; hence, many ghosts are present.

\subsection{Data acquisition and reduction}

The speckle noise attenuation capability of the testbed MCC was first quantified without the HD. These measurements were made by acquiring a sequence of images in the same bandpass for both channels by rotating the filter mosaic by 90$^{\circ}$, which made both channels fall on the same side of the mosaic. This eliminates chromatic effects and ensures that any decorrelation between the PSFs of the two channels is induced by differential aberrations. Then we obtained a sequence of images in the same configuration but with the HD in place. This sequence was obtained only a few hours after the first one, the only difference being the presence of the HD. It thus provides a direct measurement of the improvement in speckle noise attenuation provided by the HD. Finally, a third sequence was acquired with the HD using different bandpasses for the two channels, which is more representative of astronomical observations. Two diffusers (1\degr\ and 5\degr\ FWHM) from Physical Optics Corporation were tried and no significant difference in speckle attenuation was found between them; however, the 1$^{\circ}$ diffuser barely fills up the MCC pupil stop and yields a transmission of 85\% while the 5$^{\circ}$ one largely overfills the MCC pupil stop and yields a transmission of 30\%.

All sequences of images were acquired in a similar fashion. Five images with the illumination source ON and OFF were alternately obtained for a total of 40-50 PSF images and darks. Each image consisted of $\sim$5 minutes of coadditions. The intensity of the illumination source was adjusted to saturate the PSF center out to $\sim$3-4 $\lambda/D$ to improve the signal-to-noise ratio of the speckles at large separations.

Basic image reduction consisted in dark subtraction and division by the flat field. Bad pixels were replaced by the median value of neighboring pixels. For each sequence of images, the center of the first image was found by cross-correlation with a theoretical PSF having the same FWHM in an annulus where the image is not saturated. Then the other images of the sequence (including images of the second channel) were registered to the first one by cross-correlation in the same annulus. The image areas affected by ghost artifacts were masked out for cross-correlation and noise calculation purposes. When images were acquired in two bandpasses simultaneously, the short channel images were spatially scaled up to compensate for the change in diffraction scale with wavelength; the scaling factor was found by cross-correlation. In performing the subtraction, the second channel PSF intensity was normalized by the ratio of the radial profiles of the two channels. For each sequence, all images and all differences were finally median combined.

\subsection{Attenuation results}

\begin{figure}
\epsscale{1.}
\plotone{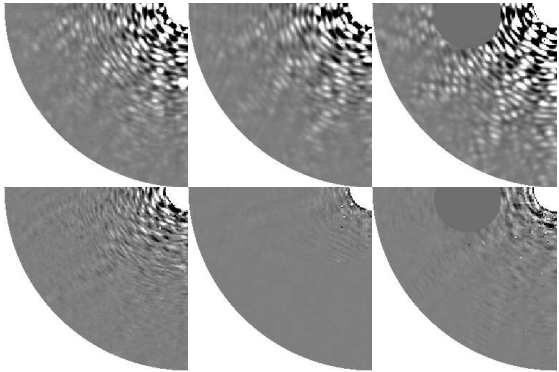}
\caption{\label{fig:im} Comparison of the image of the first channel after subtraction of a radial profile ({\it top row}) with the difference of the two channels ({\it bottom row}). The configurations are: single bandpass without HD ({\it left}), single bandpass with HD ({\it middle}), and two-bandpass with HD ({\it right}). Display intensity range is $\pm2\times10^{-5}$ of PSF peak for all images. The center of the images (white) is saturated. Regions affected by the presence of a ghost are masked out.} 
\end{figure}

\begin{figure}
\epsscale{1.}
\plotone{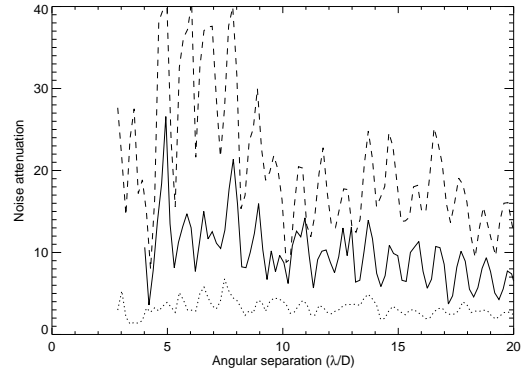}
\caption{\label{fig:att} Speckle noise attenuation: single bandpass without HD ({\it dotted line}), single bandpass with HD ({\it dashed line}), and two-bandpass with HD ({\it solid line}).}
\end{figure}

A comparison of the images before and after subtraction is shown in Figure~\ref{fig:im} and the corresponding attenuation curves are shown in Figure~\ref{fig:att}. The single bandpass attenuation without HD is $\sim$3.5 on average; this is slightly better than, but of the order of what is achieved with TRIDENT and SDI. The addition of the HD boosts the attenuation to $\sim$25, and even $\sim$40 on Airy rings, an improvement by a factor $\sim$7-10 over the configuration without the HD. This is a clear demonstration that the effect of differential aberrations is indeed reduced by using an HD. Even though the residual noise in this case is dominated by pixel-to-pixel noise, some residual speckles are present. The attenuation achieved with our testbed may be limited by a combination of interpolation errors (when shifting the images), flat-field errors, differential pixel cross-talk, differential persistence, slight optical distortion of the MCC, or differential aberrations. Misalignment of the cube and prism may also result in slight field rotation and focus differences between the two channels.

Using the HD and two bandpasses, the speckle noise attenuation achieved is $\sim$12-14. For the two wavelengths of our filter mosaic, the maximum attenuation possible for identically shaped bandpasses is $\lambda/\Delta\lambda\sim$30. However, the transmission profiles of our filters have slopes in opposite directions; this decorrelates the two PSFs further. When taking this into consideration, the maximum attenuation attainable with our mosaic is $\sim$25. Our results show an attenuation lower than this. One possible explanation for the difference, in addition to the ones mentioned previously, is an inadequate calibration of the structures produced by the irregular surface of the HD. During our measurements, we noticed a slow drift of the detector with respect to the pinhole as a function of time. This drift is likely correlated with the level of liquid nitrogen in the cryostat. Hence, the flat field could not be acquired in exactly the same configuration as the PSF images and slight calibration errors may be present. Such residuals, in the form of arcs due to the rotation of the HD, are noticeable in the residual image. The calibration of the surface of the HD is important only when different bandpasses are used because the images have to be rescaled, i.e. a given structure produced by an irregularity on the HD surface will not spatially coincide with itself after rescaling. For this reason, the performance of the two-bandpass configuration can be regarded as a lower limit; a larger attenuation could be achieved using a more stable setup.

\section{Projected on-sky performance}\label{sect:perf}

\subsection{Detection limit}

Numerical simulations were done to estimate the detection limits that currently could be achieved by a dual channel camera under three possible scenarios: (1) without an HD; (2) with an HD added to an existing MCC -- requiring the addition of a focal enlarger (\S\ref{sect:concept}) that reduces the field of view; and (3) with an MCC designed with an HD ab initio -- avoiding the focal enlarger and the reduction of the field of view. For case (2), a 4$\times$ magnification was assumed, yielding $(f/\#)_{\rm MCC}/(f/\#)_{\rm inc}=4$ as in the laboratory experiment. The baseline camera was assumed to be similar to NICI, i.e. featuring a 1024$\times$1024 pixels detector with a pixel scale of 0\farcs018 and installed on an 8-m telescope equipped with an 85-actuator AO system. The simulation was done for a wavelength of 1.587 $\mu$m, the central wavelength of the first differential imaging filter of NICI. Coronagraphy was not considered as it does not provide significant gain in contrast for Strehl ratios typically achieved today, i.e. $<$50\% \citep{sivaramakrishnan01}.

A pupil of diameter 456 pixels placed at the center of a 2048$\times$2048 pixels array was used to produce a PSF with a FWHM equal to 4.5 pixels; the corresponding pixel scale is 0\farcs009 at the wavelength of the simulation. Atmospheric phase screens were generated for $r_0=20$~cm at 0.5~$\mu$m. Additionally, a static wavefront phase error having 80~nm rms and a power-law power spectrum of index -2.7 was used. AO phase filtering was simulated by multiplying the amplitude of the phase Fourier transform by a high-pass filter equal to $(k/k_{\rm AO})^n$ for $k<k_{\rm AO}$ and to 1 for $k\geq k_{\rm AO}$, where $k$ is the spatial frequency of the phase aberration and $k_{\rm AO}$ is the AO cutoff frequency \citep{sivaramakrishnan01}. The filter index $n$ was set to 1.55 to produce an average Strehl ratio of 0.3, the baseline value expected for NICI\footnote{see the NICI Campaign Science documentation at\\ http://www.gemini.edu/sciops/instruments/nici/niciCampaign\_orig.html.}. A long exposure monochromatic PSF was obtained by coadding the result of 5000 independent atmospheric wavefront realizations. A polychromatic PSF image was then obtained by coadding 101 spatially scaled versions of the monochromatic image, each representing a different wavelength within the 1\% bandwidth of the filter. This PSF image was then either spatially scaled up by a factor of 2, or its pixels binned $2\times2$ to obtain proper sampling for the configuration with the 4$\times$ magnification module (0\farcs0045/pixel), or without this module (0\farcs018/pixel), respectively. The speckle noise profiles were computed from these images.

\begin{figure}
\epsscale{1.}
\plotone{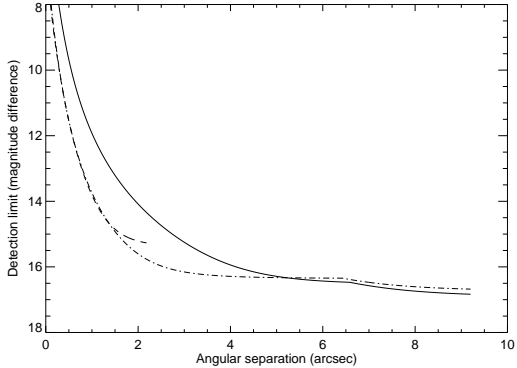}
\caption{\label{fig:detlim} Detection limit in difference of magnitude for an $H=6$ star and exposure time of 3600~s without HD ({\it solid line}), with HD and 4$\times$ focal enlarger ({\it dashed line}), and with HD and no focal enlarger ({\it dot-dashed line}). The curve for the configuration with the focal enlarger is shorter due to the reduced field of view.}
\end{figure}

The residual speckle noise was then obtained by assuming attenuation factors of 2.5 and 14 for the configuration without and with HD respectively. The pixel-to-pixel noise was computed for a star of given $H$-band magnitude assuming a total system efficiency of 24\%\footnote{This is the estimated total system efficiency of NICI, \\see http://www.gemini.edu/sciops/instruments/nici/niciCampaign\_orig.html.} and 19\% per channel without and with the HD respectively, a sky background of 14 mag/arcsec$^2$ in $H$,  a dark current of 0.15 e$^-$/s per pixel, 30 exposures of 120 s and a read noise of 10 e$^-$/pixel per exposure. Since the statistical distribution of static speckles is non Gaussian, a 5$\sigma$ detection limit is not appropriate; rather, to reach a confidence level similar to that of a 5$\sigma$ Gaussian confidence interval, the detection limits were obtained as the quadratic sum of 10$\sigma_{\rm speckle}$ and 5$\sigma_{\rm pixel}$ (C. Marois et al., in preparation), where $\sigma_{\rm speckle}$ is the speckle noise (dominating at small separations) and $\sigma_{\rm pixel}$ is the pixel-to-pixel noise (dominating at large separations). A correction was applied to these limits to account for the partial loss in planet flux from the two-channel subtraction\footnote{A maximum loss of 12\% was assumed; this is the expected flux loss for a T8 dwarf ($T_{\rm eff} \sim 800$~K) using the narrow band filters of NICI. According to the models of \citet{baraffe03}, giant planets of 5 M$_{\rm Jup}$ or less and older than a few 10 Myr should have $T_{\rm eff}<800$~K and thus methane absorption redward of 1.6 $\mu$m is expected to be at least as important for such objects as for T8 dwarfs.}; this loss is maximal at zero separation and decreases to zero at a separation where the spatial scaling of the images displaces sufficiently (by $\sim$1 $\lambda/D$) the planet. The results for an $H=6$ star are shown in Figure~\ref{fig:detlim}. The higher asymptotic limit for the HD with 4$\times$ magnification is due to the higher relative importance of read noise for this largely oversampled configuration. The slightly higher asymptotic limit of the HD without magnification compared to the configuration without HD is explained by the 20\% loss in throughput caused by the HD.

The detection limits presented in Figure~\ref{fig:detlim} should not be regarded as predictions of the best contrast that a dual-channel camera could reach in practice since other speckle suppression strategies, such as angular differential imaging or subtraction of images obtained after rotation of the instrument, could be used to attenuate the noise further. We note however that the detection limit without HD presented in this figure is very similar to the estimated baseline performance of NICI\footnote{see the NICI Campaign Science documentation at\\ http://www.gemini.edu/sciops/instruments/nici/niciCampaign\_orig.html. When comparing both curves, one should keep in mind that the detection limits of Figure~\ref{fig:detlim} assume a 10$\sigma_{\rm speckle}$ criterion whereas the curve in the NICI documentation is for 5$\sigma_{\rm speckle}$, this translates into a difference of 0.75~mag.}.

\subsection{Detection efficiency}

A Monte Carlo simulation was done to investigate the impact of using an HD in terms of actual planet detections. First, realistic target samples were constructed for three representative populations of interest for planet searches. In particular, the distance, the age, and the $H$-band magnitude of each star of the samples were determined. The three samples are:

\begin{enumerate}

\item NYS $-$ Nearby young stars ($\lesssim 150$~Myr, $<50$~pc). This sample includes 72 stars at a distance less than 50~pc from the Sun, obtained from Tables 3 and 4 of \citet{wichmann03}; these stars have ages similar to those of the Pleiades stars ($20-150$~Myr) based on lithium abundance measurements.  The stellar distances were taken from the same reference. Since the precise age of most of these stars is not well known, the stars were randomly assigned an age from a Gaussian distribution of mean 2.0 and standard deviation 0.15 in $\log$(Age/Myr).

\item TWA $-$ TW Hydrae Association ($\sim$8 Myr, 30-60 pc). This sample includes 23 stars from Table~1 in \citet{zuckermann04}. The stellar distances were taken from the same reference. The stars were randomly assigned an age from a Gaussian distribution of mean 0.9 and standard deviation 0.04 in $\log$(Age/Myr).

\item $\rho$~Oph $-$ $\rho$~Ophiuchus star forming cloud ($\sim$2 Myr, $\sim$135 pc). This sample includes 27 stars with $R<15$ from Table~4 of \citet{wilking05}. The ages were taken from the same reference. The stellar distances were uniformly sampled in the 120-150~pc interval \citep{knude98}.

\end{enumerate}

\begin{figure}
\epsscale{1.15}
\plotone{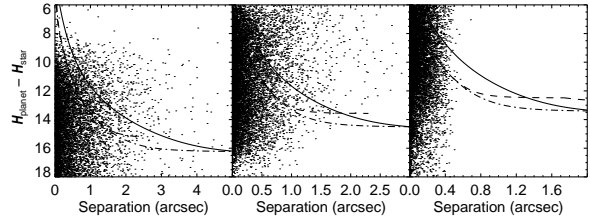}
\caption{\label{fig:mc} Result of the Monte Carlo simulation for the NYS ({\it left}), TWA ({\it middle}), and $\rho$~Oph ({\it right}) samples. The curves show the detection limits for a star having an $H$ magnitude equal to the median of the sample for the configuration without HD ({\it solid line}), with HD and 4$\times$ magnification unit ({\it dashed line}), and with HD and no magnification unit ({\it dot-dashed line}). On the left and middle panels, the curve for the configuration with the focal enlarger is shorter due to the reduced field of view.}
\end{figure}

In all cases the $H$-band magnitudes were taken from the 2MASS point source catalog (PSC, \citealp{cutri03}). For the simulation purpose, all samples were artificially increased to a total of 50000 targets by re-using the same stars with different values for the random parameter. A population of exoplanets (one planet per star) was then generated using a mass distribution following $\drm N/\drm M \propto M^{-1}$ with $0.5 < M/{\rm M}_{\rm Jup} < 12$, a semi-major axis distribution following $\drm N/\drm a \propto a^{-0.5}$ with $0.1 < a/{\rm AU} < 50$, and a Gaussian eccentricity distribution of mean 0.25, standard deviation 0.19 and with $0 \leq e \leq 0.8$; these properties were taken from \citet{marcy05} except for $\drm N/\drm a$ which was set according to a minimum mass solar nebula. All distributions are consistent with the exoplanet population from radial velocity surveys. The orbital inclination and phase were obtained by uniformly sampling respectively the orbital orientation and the orbital area swept by the planet. Using the stellar distances and the orbital parameters, the apparent separation in arcsec was computed for all planets. Models of \citet{baraffe03} were used to derive the apparent $H$-band magnitude of the planets given their mass, age and distance. The magnitude difference between the planets and their primary star was finally compared to the detection limits. The results are shown in Figure~\ref{fig:mc} and the corresponding detection efficiencies are presented in Table~\ref{tbl}.
The detection efficiencies for the configuration without an HD are respectively 6\%, 17.5\%, and 12.5\% for the NYS, TWA, and $\rho$~Oph samples respectively. The use of an HD roughly doubles these numbers. In all cases it is interesting to note that one would not suffer from the reduction in FOV needed to use an HD with an existing camera, the reason being that most planets are located at small angular separations. Given these relatively small planet detection efficiencies, the relatively small number of stars in real target samples, and the unknown but likely small fraction of stars with giant planets on large orbits, it is reasonable to expect that a dedicated search for planets using actual AO systems on 8-m telescopes will uncover at most a handful of planets. In this case, i.e. for small number statistics, the factor of two in detection efficiency resulting from the use of an HD could make a significant difference.

\begin{deluxetable}{lccc}
\tablecolumns{4}
\tablewidth{170pt}
\tablecaption{Planet detection efficiency \label{tbl}}
\tablehead{
\colhead{Configuration} & \multicolumn{3}{c}{Target sample} \\
\cline{2-4}
  & \colhead{NYS} & \colhead{TWA} & \colhead{$\rho$~Oph}
}
\startdata
No HD                & 0.063 & 0.175 & 0.125 \\
HD w/ 4$\times$ mag. & 0.113 & 0.299 & 0.256 \\
HD w/o mag.          & 0.128 & 0.295 & 0.244
\enddata
\end{deluxetable}

\section{Discussion}\label{sect:discussion}

The experiment presented in this paper validates the concept of breaking the wavefront coherence before separating it into multiple wavelengths for SSDI. It is clearly demonstrated that this strategy reduces the limitations imposed by differential aberrations. Still, the two-bandpass attenuation obtained with our testbed could have been as high as 25 considering the wavelengths of the filter mosaic used. We do not believe that the attenuation achieved was limited by differential aberrations, nor that it represents a fundamental limit to the attenuation achievable with such a system. The fact that the single-bandpass attenuation was higher offers evidence for this claim. We believe that a dedicated MCC built with more care, i.e. with baffles to reduce scattered light, with coatings on all surfaces to reduce ghost artifacts, with a better alignment of the prism with respect to the beam splitter to minimize differential rotation and distortion between channels, with a more precise control of co-focality between channels, and with better mechanical stability, would provide better noise attenuation. The design of NICI correctly addresses these issues and thus NICI should not suffer from these shortcomings. For completeness, we estimated how the planet detection efficiency would improve with better speckle noise attenuation. It was previously mentioned that the maximum attenuation is $\sim$30 for bandpasses appropriate for exoplanet searches. With such an attenuation, the planet detection efficiency would increase further by $\sim$25\%.

The experimental validation of the concept that we presented also offers a good indication that an integral field spectrograph based on a microlens array (e.g. GPI, \citealp{macintosh_gpi}; SPHERE/IFS, \citealp{sphere}; or PFI, \citealp{macintosh_pfi}) should work well. In such a system, a microlens array is placed at the focal plane to spatially sample the PSF. Each microlens produces a small pupil that is dispersed and re-imaged onto a detector by a spectrograph; a spectrum of each spatial sample of the PSF is thus obtained. Data processing allows the construction of a PSF data cube, containing a spectrum for each spatial pixel. Similarly to the HD, the spatial sampling of the PSF made by the microlens array breaks the coherence of the wavefront and each ``micro pupil'' becomes an independent source for the spectrograph. In this case, differential aberrations in the spectrograph affect only the light distribution within the spectra and not the speckles of the PSF imaged. Hence, images of the PSF at different wavelengths should be highly correlated. Preliminary laboratory tests of an integral field spectrograph prototype seem to corroborate this hypothesis \citep{lavigne06}. Based on a similar argument, the use of a microlens array at the entrance focal plane of an MCC, a concept presented in \citet{dl_spie2004}, should be a good alternative solution.

It was previously mentioned that an existing MCC could be adapted for use with an HD, we take NICI as an example and briefly state the type of work that would be needed to do so. The NICI system (see \citet{nici})\footnote{see also http://www.gemini.edu/sciops/instruments/nici/niciIndex.html} consists of a complete AO system followed by a dual-channel camera operating in the 1-5~$\mu$m wavelength interval. The focal plane of the camera is located outside of the cryostat. At this focal plane, a wheel mechanism is used to select between various occulting masks for coronagraphy. To implement an HD in this system it would thus be necessary to replace the occulting mask wheel by a mechanism to hold the HD and make it rotate or move continuously as explained in Sect.~\ref{sect:concept}. This mechanism would not be technically challenging to realize as it would be warm. The entrance focal ratio of the camera $(f/\#)_{\rm MCC}$ is 16, thus a diffuser diverging light within a cone of half-angle equal to 1.8$^\circ$ would be needed; a custom HD matched to this angle should be manufacturable as this is in the range of angles available from Physical Optics Corporation. As explained previously, it would be necessary to enlarge the exit focal ratio of the AO system ($f/16$); based on our numerical simulations and experimental results, a factor of 3-4 appears adequate. This magnification would be provided by a system of lenses placed between the dichroic of the AO system and the HD mechanism; these lenses would be warm. Accordingly, the field of view would be reduced by the same factor. These modifications would preclude the use of coronagraphy and would increase the thermal background at longer wavelengths. However, the latter is irrelevant for planet searches as SSDI observations are made in the $H$-band, and the loss of coronagraphic capabilities would be largely over compensated by the increase in speckle noise suppression given the modest Strehl ratios expected.

\section{Conclusion}

The concept of placing an HD at the entrance focal plane of an MCC to minimize the impact of differential aberrations in SSDI was presented. This concept was shown experimentally to improve significantly the speckle noise attenuation: our dual-beam dual-bandpass experimental testbed produced a speckle noise attenuation of $\sim$12-14, a factor of $\sim$5 improvement over existing SSDI cameras. Using Monte Carlo simulations of realistic exoplanet populations, it was shown that such an improvement in speckle noise attenuation could yield an increase by a factor $\sim$2 of the number of potentially detectable exoplanets. The option presented in this paper should be considered seriously for upcoming multi-channel cameras and instruments dedicated to exoplanet detection.

\acknowledgments

The authors would like to thank Ren\'e Racine for helpful discussions, and Philippe Vall\'ee, Todd Szarlan, Jean-S\'ebastien Mayer and Guillaume Provencher for their help in designing and machining parts used for the experimental setup. The authors would also like to thank the anonymous referee for suggestions that improved the quality of the paper. This work was supported in part through grants from the Natural Sciences and Engineering Research Council, Canada, and from the Fonds Qu\'eb\'ecois de Recherche sur la Nature et les Technologies.

\begin{appendix}

\section{VLT SDI two-channel subtraction attenuation} \label{sect:appendix}

\begin{figure}[h!]
\epsscale{0.6}
\plotone{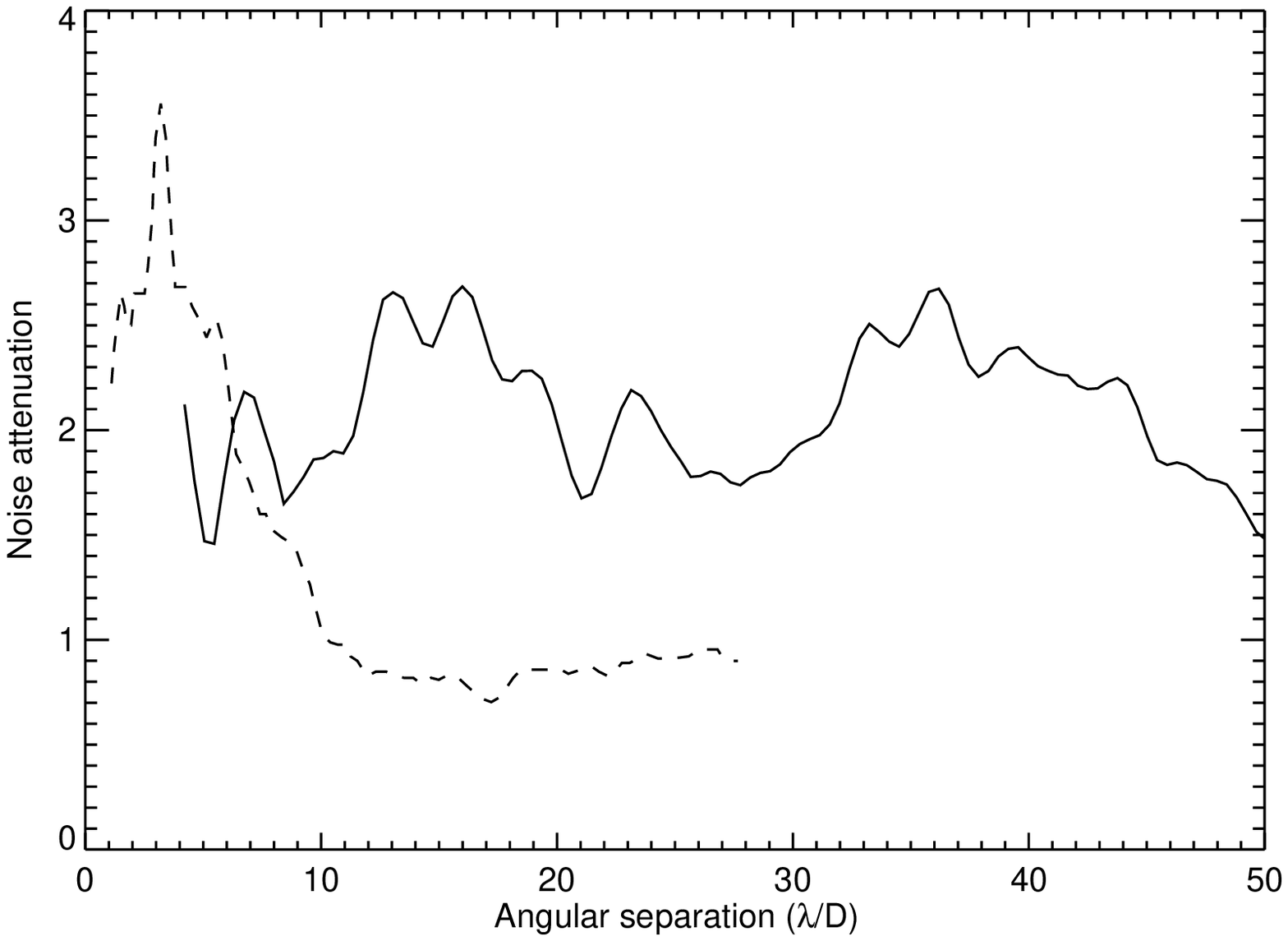}
\caption{\label{fig:sdi} Noise attenuation resulting from the subtraction of the two 1.625~$\mu$m images with SDI on the VLT ({\it solid line}). For completeness, the noise attenuation resulting from the subtraction of the 1.58~$\mu$m and 1.625~$\mu$m images with TRIDENT at the CFHT is shown as a dashed line; these observations were read noise limited beyond 10 $\lambda/D$ (curve adapted from \citealp{trident}).}
\end{figure}

Since the speckle noise attenuation resulting from a two-channel subtraction of images using SDI is unavailable in the literature, it is estimated here\footnote{Based on observations made with ESO Telescopes at the Paranal Observatories under program ID 60.A-9026.}. Observations of the star AB Dor obtained on 2004 February 1 \citep{close05} were retrieved from the European Southern Observatory science archive and used to quantify the performance of SDI. This dataset consists in two sets of images at five dither positions obtained at position angles differing by 33 degrees. SDI produces four simultaneous PSF images: one at 1.575 $\mu$m, one at 1.600 $\mu$m, and two at 1.625 $\mu$m. The reduction steps consisted in the subtraction of a sky frame, division by a flat field image and correction of bad pixels using the median of neighboring pixels. The PSF images of all channels were then spatially scaled to match the 1.625 $\mu$m channel and registered to a common center with subpixel accuracy. When two PSF images were subtracted, their intensity was adjusted to minimize the residual noise. The six possible two-channel subtractions were performed and a noise attenuation $\sim$2.-2.5 was obtained for all cases. Figure~\ref{fig:sdi} shows the mean attenuation curve over the ten images for the subtraction of the two 1.625 $\mu$m channels. We stress that this attenuation is not the best achievable with the SDI device, subtraction of simultaneous multi-channel difference images obtained at different instrument rotation angles have been shown to attenuate further the residuals; see \citet{biller04} for more detail.

\end{appendix}

\end{document}